# Two-Pion Exchange Currents in Photodisintegration of the Deuteron

D. Rozpędzik[a], J. Golak[a], S. Kölling[b,c] and E. Epelbaum[d]

[a] *M. Smoluchowski Institute of Physics, Jagiellonian University, PL-30059 Kraków, Poland*
[b] *Forschungszentrum Jülich, Institut für Kernphysik (IKP-3), D-52425 Jülich, Germany, and Jülich Center for Hadron Physics, D-52425 Jülich, Germany*
[c] *Helmholtz-Institut für Strahlen- und Kernphysik (Theorie) and Bethe Center for Theoretical Physics, Universität Bonn, D-53115 Bonn, Germany*
[d] *Institut für Theoretische Physik II, Ruhr-Universität Bochum, D-44780 Bochum, Germany*

**Abstract.** Chiral effective field theory (ChEFT) is a modern framework to analyze the properties of few-nucleon systems at low energies [1]. It is based on the most general effective Lagrangian for pions and nucleons consistent with the chiral symmetry of QCD. For energies below the pion-production threshold it is possible to eliminate the pionic degrees of freedom and derive nuclear potentials and nuclear current operators solely in terms of the nucleonic degrees of freedom. This is very important because, despite a lot of experience gained in the past, the consistency between two-nucleon forces, many-nucleon forces and the corresponding current operators has not been achieved yet. In this presentation we consider the recently derived long-range two-pion exchange (TPE) contributions to the nuclear current operator which appear at next-to leading order of the chiral expansion [2]. These operators do not contain any free parameters. We study their role in the deuteron photodisintegration reaction and compare our predictions with experimental data from Refs [3]. The bound and scattering states are calculated using five different chiral N2LO nucleon-nucleon (NN) potentials which allows to estimate the theoretical uncertainty at a given order in the chiral expansion. For some observables the results are very close to the reference predictions based on the AV18 NN potential and the current operator (partly) consistent with this force.

## INTRODUCTION

Chiral effective field theory (ChEFT) provides a systematic and model-independent framework to analyze hadron structure and dynamics in harmony with the spontaneously broken approximate chiral symmetry of QCD. This approach is a powerful tool for the derivation of the nuclear forces. Exchange vector and axial currents in nuclei have also been studied in the framework of ChEFT. Since the pioneering work of Park et al. [4], heavy-baryon chiral perturbation theory has been applied to derive exchange axial and vector currents for small values of the photon momentum. These calculations were carried out in time-ordered perturbation theory. The resulting exchange vector currents were, in particular, applied to analyze radiative neutron-proton capture within a hybrid approach.

ChEFT has been also used to study the electromagnetic properties of the deuteron [5]. One of the most fundamental processes on the deuteron is the photodisintegration reaction. It has been a subject of intensive experimental and theoretical research for several decades (see Refs [6]). However, very few applications to this reaction at small momentum transfer have been carried out so far. A recent review on the theoretical achievements in this field based on conventional framework can be found in Ref. [7]. A strong interest for this reaction within ChEFT, especially in view of planned experiments, resulted in the application of chiral framework. This requires a consistent derivation of the nuclear Hamiltonian and the electromagnetic current operator for a given few-nucleon system.

In the two-nucleon (2N) system, the leading contributions to the exchange current originate from one-pion exchange and are well known. The corrections to the 2N current operator at the leading one loop level in ChEFT were recently worked out by Pastore et al. in Ref. [8] based on time-ordered perturbation theory. Independently, the two-pion exchange 2N current operator was derived in Ref. [2] using the method of unitary transformation. The resulting current operator is consistent with the corresponding chiral two-nucleon potential [1] obtained within the same scheme. In the present work, we explore the effects of the leading two-pion exchange 2N operator [2] in the deuteron photodisintegration reaction. We emphasize, however, that the presented calculation is not yet complete. In particular, the corresponding expressions for the one-pion exchange and short-range contributions to the current operator within the method of unitary transformation are not yet available. The important outcome of our investigation is that various observables in the deuteron photodisintegration appear to be highly sensitive to the two-pion exchange current.

## FORMALISM

The formalism to describe 2N electromagnetic reactions requires the knowledge of the consistent potential and electromagnetic current. The NN potential based on ChEFT is well known up to next-to-next-to-next-to-leading order in the chiral expansion. As already pointed out, we restrict ourselves in this study to next-to-leading order (NLO) long-range contributions to the 2N current operator. To achieve a better accuracy, all calculations are made using the N2LO potential. At this order, it contains the one-pion-exchange (OPE) ($1\pi$) and two-pion exchanges (TPE) ($2\pi$) contributions as well as various contact interactions (cont) [1]

$$V_{2N} = V_{1\pi} + V_{2\pi} + V_{cont}. \tag{1}$$

The effective current operator for the 2N system is a sum of the single-nucleon operators $J^\mu(i)$, $i=1,2$ and two-nucleon operators of different type ($J^\mu(1,2)$)

$$J^\mu_{2N} = J^\mu(1) + J^\mu(2) + J^\mu_{1\pi}(1,2) + J^\mu_{2\pi}(1,2) + J^\mu_{cont}(1,2), \tag{2}$$

where the expressions for the single-nucleon and OPE currents are well established.

In this contribution, we concentrate on a treatment of the long-range TPE contributions to the 2N current operator derived in Ref. [2]. The expressions for the TPE current operator in momentum space are rather involved and contain the standard loop functions and the three-point functions in a form suitable for numerical calculations.

The TPE four-current operator $J^\mu = (J^0, \vec{J})$ can be decomposed according to its isospin-spin-momentum structure and quite generally written in the form

$$J^0 = \sum_{\alpha=1}^{5}\sum_{\beta=1}^{8} f_\alpha^{\beta S}(\vec{q}_1, \vec{q}_2) T_\alpha O_\beta^S \quad ; \quad \vec{J} = \sum_{\alpha=1}^{5}\sum_{\beta=1}^{24} f_\alpha^\beta(\vec{q}_1, \vec{q}_2) T_\alpha \vec{O}_\beta, \tag{3}$$

where $\vec{q}_i \equiv \vec{p}' - \vec{p}$ is the momentum transferred to nucleon $i$, $T_\alpha$ is the 2N isospin operator, $O_\beta^S$ and $\vec{O}_\beta$ are the (momentum dependent) spin operators in the 2N space, $f_\alpha^{\beta S}$ and $f_\alpha^\beta$ are scalar functions. The explicit form of the scalar functions can be found in Ref. [2]. The expressions do not contain any free parameters. Here and in what follows, we use the operator basis for $O_\beta^S$ and $\vec{O}_\beta$ given in Ref. [2].

Due to their isospin structure, not all combinations of (3) contribute to photodisintegration of the deuteron. The non-vanishing contributions emerge from

$$\vec{J} = \sum_{\beta=3}^{10} f_2^\beta(\vec{q}_1, \vec{q}_2) T_2 \vec{O}_\beta + f_3^2(\vec{q}_1, \vec{q}_2) T_3 \vec{O}_2. \tag{4}$$

Further information about these operators can be found in Ref. [2].

We choose a chiral potential $V_{2N}$ to generate the deuteron bound state, $|\psi_{bound}\rangle$, and the proton-neutron (pn) scattering state $|\psi_{scatt}^{pn}\rangle$. These are needed in order to obtain the nuclear matrix element $N^\mu$

$$N^\mu = \langle \psi_{scatt}^{pn} | J^\mu_{2N} | \psi_{bound} \rangle, \tag{5}$$

from which all observables can be calculated. We use the solution of the Lippmann-Schwinger equation, $t = V_{2N} + tG_0V_{2N}$, in order to express $N^\mu$ as

$$N^\mu = \langle \vec{p}_0 | (1+tG_0) J^\mu_{2N} | \psi_{bound} \rangle, \tag{6}$$

where $G_0$ is the free 2N propagator and $\vec{p}_0$ is the relative pn momentum in the final state.

We work in momentum space and employ the standard partial wave decomposition of the chiral potential, see for example Ref. [9]. Thus we have to express the TPE current operator in the same partial wave basis. To this end, we first prepare all the spin and isospin matrix elements using *Mathematica©* and then calculate the resulting four-fold angular integrals (7) on the parallel supercomputer IBM Blue Gene/P of the Jülich Supercomputing Centre (JSC).

$$\langle p'(l's')j'm';t'm_{t'}|\vec{J}_{\alpha\beta}|p(ls)jm;tm_t\rangle = \int d\hat{p}' \int d\hat{p} \sum_{m_{l'}} \sum_{m_l} C(l's'j';m_{l'},m'-m_{l'},m') Y^*_{l'm_{l'}}(\hat{p}')$$
$$\times C(lsj;m_l,m-m_l,m) Y_{lm_l}(\hat{p}) f^\beta_\alpha(\vec{q}_1,\vec{q}_2) \langle t'm_{t'}|T_\alpha|tm_t\rangle \langle s'm'-m_{l'}|\vec{O}_\beta|sm-m_l\rangle. \quad (7)$$

## RESULTS

We now discuss the results for the deuteron photodisintegration process for the unpolarized cross section and selected polarization observables. The results for the differential cross section and the photon analyzing power at the photon laboratory energies of $E_\gamma=$ 30, 60, 100 MeV are shown in Figure 1. Note that the theoretical predictions depend on the two cut-off parameters that appear in the chiral potential. While the first cut-off parameter $\Lambda$ appears in the regulator function for the Lippmann-Schwinger equation, the second parameter $\tilde{\Lambda}$ enters the spectral function regularization (SFR) and denotes the ultraviolet cut-off value in the mass spectrum of the two-pion-exchange potential. The bands reflect the dependence of the results on variations of $\Lambda$ and $\tilde{\Lambda}$. Following Ref. [1], the cut-off values are varied between 450 and 600 MeV for $\Lambda$ and between 500 and 700 MeV for $\tilde{\Lambda}$. The plots show the contributions from the different parts of the 2N current: single-nucleon current (blue band), one-pion exchange contribution (dark band) and the long-range TPE contributions (pink band). As a reference, we also show the results based on the phenomenological AV18 potential [10] and the corresponding exchange currents. We have also calculated the deuteron tensor analyzing powers as a function of the proton emission angle for three $E_\gamma$- energy bins. These results are visualized in Figure 2.

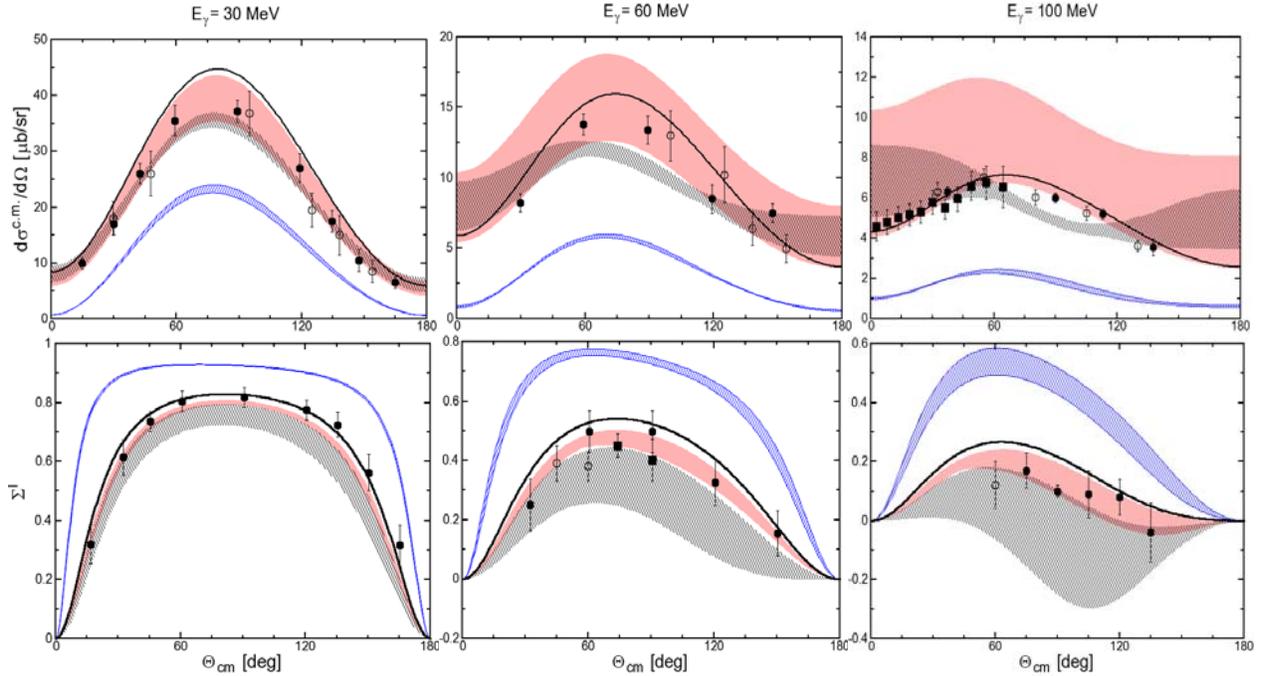

**FIGURE 1.** (color online) Results for the unpolarized cross section and the photon analyzing power in the deuteron photo-disintegration process at the photon laboratory energies of $E_\gamma=$ 30, 60, 100 MeV, displayed as functions of the proton emission angle. The solid black line refers to the standard calculation based on the AV18 potential, the blue band covers results obtained with the single-nucleon current only, the dark band represents the predictions based on the single-nucleon and OPE parts and the pink band includes, in addition, the contributions of the TPE current. The experimental data are from Ying et al. [3].

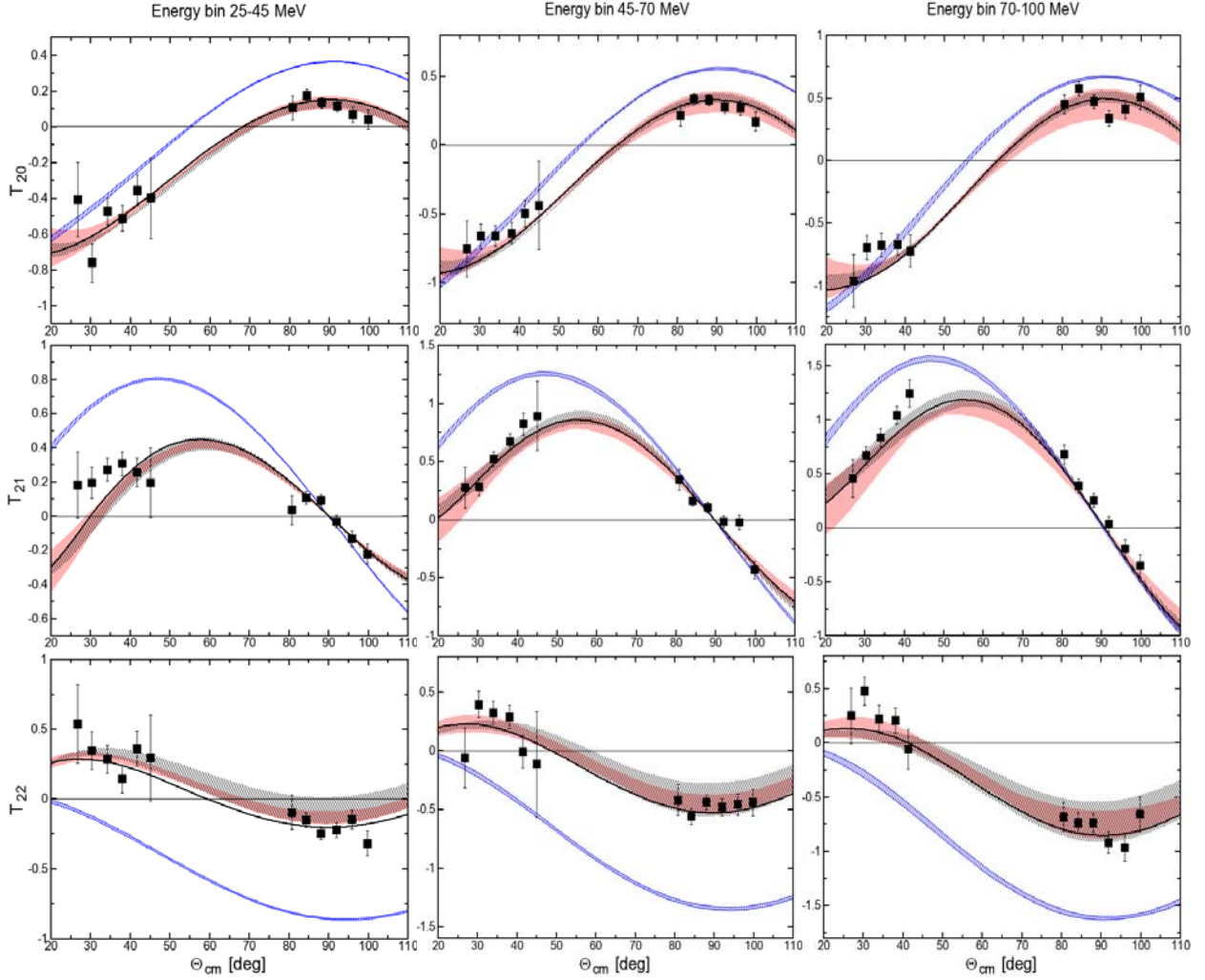

**FIGURE 2.** (color online) Deuteron tensor analyzing powers vs. proton emission angle for three $E_\gamma$- energy bins. The bands and lines have the same meaning as in Figure 1. The theoretical predictions are compared to the experimental data from Rachek et al. [3]. The pictures do not show systematic errors.

We see that in all observables the 2N current operator plays an important role and a restriction to the single nucleon current operator only leads to a strong disagreement with the data. One also observes a rather good agreement between the calculations based on the chiral potential/current operator and the ones based on the AV18 potential. In general, our results agree reasonably well with the experimental data. Definite conclusions can, however, only be drawn after carrying out a complete calculation including *all* ingredients present at this chiral order. This work is in progress. We also expect that such a complete calculation will yield a much narrower bands as the cut-off dependence should, to a large extent, be absorbed into the "running" of the corresponding short-range current operators.

## CONCLUSIONS

We explored the effects of the TPE currents derived recently in the framework of ChEFT in the deuteron photodisintegration reaction. We also studied the effects of various other ingredients of the chiral 2N current operator in the unpolarized cross section and several polarization observables. As a main outcome of our study, we find that the new terms in the exchange current operator beyond the well-known one-pion exchange contribution play an important role in this reaction. Our dynamical framework is still in development. The inclusion of the consistent short-range contact and higher-order OPE contributions to the 2N current operator is in progress. We also plan to extend these calculations to the three-nucleon system. Finally, we emphasize that new, hight-quality data will be necessary to fix unknown parameters in the short-range part of the exchange current and to test the predictions of ChEFT.

This work was supported in part by the Polish Ministry of Science and Higher Education (grants N N202 104536, N N202 077435).

## REFERENCES

1. E. Epelbaum, *Prog.Part. Nucl. Phys.*, **57**, 654 (2006) and references therein.
2. S. Kölling et al., *Phys. Rev. C,* **80**, 045502 (2009).
3. S. Ying et al. *Phys. Rev. C,* **38***,* 1584-1600 (1988)*;* I. A. Rachek et al., *Phys. Rev. Lett.,* **98**, 182303 (2007).
4. T. S. Park et al. *Phys. Rept*., **233**, 341(1993).
5. M. Walzl and U.-G. Meissner, *Phys. Lett. B*, **513**, 37 (2001); D. Phillips, *PoS CD*, **09**, 066 (2009).
6. H. Arenhövel et al., *Few-Body Sys. Suppl.* **3**, 1 (1991); R. Gilman et al., *J. Phys. G, Nucl. Part. Phys.* **28**, R37 (2002).
7. J. Golak et al. *Phys. Rept.*, **415**, 89 (2005) and references therein.
8. S. Pastore et al., *Phys. Rev. C*, **78**, 064002 (2008).
9. W. Glöckle et al. "The QM Few-Body Problem" (1983).
10. R. B. Wiringa et al., *Phys. Rev. C*, **51**, 38 (1995).